# FAIME:
# A Framework for AI assisted Musical Devices

**Miguel Civit[1]**, *Luis Muñoz–Saavedra[2], Francisco Jose Cuadrado[1], Charles Tijus[4] and Maria J.Escalona[3]*

1. Facultad de Ciencias Sociales y Humanas, Universidad Loyola, Seville , Spain.
2. Robotics and Computer Technology Lab, ETSII,University of Seville, Seville, Spain.
3. Computer Languages and Systems Department, ETSII, University of Seville, Seville, Spain.
4. University of Paris, France

   * Correspondence: mcivit@uloyola.es

**Featured Application: This work has direct applications to Music Performance by persons with disabilities.**

**Abstract:** *In this paper we present a novel framework for the study and design of AI assisted musical devices (AIMEs). Initially, we present a taxonomy of these devices and illustrate it with a set of scenarios and personas. Later, we propose a generic architecture for the implementation of AIMEs and present some examples from the scenarios. We show that the proposed framework and architecture are a valid tool for the study of intelligent musical devices.*

**Keywords:** Artificial Intelligence, Musical Devices, Internet of Musical Things

## 1. Introduction

Advances in technology and computer science have greatly enhanced the possibility of designing, developing, and deploying intelligent musical devices. A typical well-studied subset of these intelligent devices are IoMusTs (Internet of Musical Things). According to [1] a IoMusT is a "computing device capable of sensing and exchanging data to serve a musical purpose". A IoMusT does not need to be able to produce, select or modify music but it can be any device that is "music aware" in the sense that its behavior is directly related to music. As an example PixMob devices [2] have been widely used in musical performances. These devices that can be either worn (smartband), thrown (balls) or attached to the audience seats and are able to produce light patterns synchronized with life performances.

Not all intelligent musical devices are IoMusts. We can design intelligent devices where the intelligence is embedded in the device, and thus we may say that we gave an Intelligent Musical dEvice (IME) but not as a part of the Internet of Musical Things. In [3] we study the evolution of the design of intelligent musical instruments. In most cases, these instruments use artificial intelligence as a tool for user interaction without requiring any connection to public networks or cloud-based services. It is important to consider that machine learning ML, in most of these cases, cannot be considered as an independent agent but mainly as one of the possible alternatives for designing layers of a complete system. These types of device can also be considered as cyber-physical systems, as they clearly require intelligent software systems and dedicated hardware.

In this work, we will create a framework that covers all, or at least a very wide part, of intelligent musical devices and helps design, understand, and study them.

The rest of the paper is divided as follows: First, in the Materials and Methods section, the different datasets and published works are detailed, as well as the analysis methodology used to test the different systems. The results obtained after training and testing the different systems are then detailed and explained in the Results and Discussion sections. Finally, conclusions are presented.

## 2. Materials and Methods

Artificial intelligence-assisted musical devices come in a wide variety of forms and potentially have a very wide spectrum of uses. In order to create a framework that will cover most of these possibilities, we will start by introducing a taxonomy of the different usages of the said devices. It should be clear that it is possible for a device to fall into several categories. As an example, most musical instruments could also be considered educational aids, some of them being used predominantly for this purpose. The monochord was used through the middle ages for educational and scientific purposes [4] and similarly we can design intelligent instruments that, although being able to be used for performing, are meant with an educational intent.

*2.1. Taxonomy*

We propose a classification for AI-assisted musical devices (AIMEs). It is clear that this is not the only possible taxonomy, but it is complete, easy to apply, and useful. The classification is shown in Table 1.

In a first level, we divide our AIMEs into:

- Devices that are played by musicians: Musical instruments.
- Devices designed to modify music: Music Processors.
- Devices that compose music: Music Generators.
- Devices that select music: Music Recommenders.
- Devices that send to the user or environment information extracted from the music: Feedback systems.
- Devices designed to be used in an educational process: Educational devices.

It is clear that a real device may be included in several categories. As an example, a device could generate a set of music scores and then recommend some of them to a student. In this way, this device could be considered as a generator, a recommender, and an educational system.

This main AIME division can then be divided into subcategories. As an example, a Music Generator can either be instrumental, vocal, or combined. An instrumental music generator usually produces music in symbolic format. The most common symbolic format is the Musical Instrument Digital Interface (MIDI), which contains information that indicates the pitch, start time, stop time, and other properties of each individual note, rather than the resulting sound. Combined and voice generators have to use a raw audio format and are much more difficult to implement, although their quality has improved significantly in the present decade [5,6].

As a further example, recommendation devices can recommend music as a function of the environment or as a function of the user state. The environment-based recommendation is mostly used in social scenarios, e.g. if the system selects music for a shopping mall or an elevator. Personal Music recommendation devices are used mostly when recommending for a single user. As an example, we could estimate the user's emotional state from the data of the wearable device [7] and select the music accordingly. It is also possible to use the acquired data of an AIME personal recommender to try to modify some aspects of user behavior. An interesting possibility would be to train the user, through music, to reduce his or her stress level. In this way, the device could also be considered as part of the Internet of Behavior (IOB) [8].

*2.2. Scenarios and Personas*

In this subsection, we use Nielsen's user-centered approach [9] to define a set of scenarios and personas that would allow us to shed light on the design ideas behind different musical things. It is important to mention that devices should be evaluated in terms of functionality, usability, and user experience [10]. We will propose a set of scenarios and introduce different personas as they become relevant in relation to these scenarios. It is also clear that user sentiments have a very significant impact on the user experience with musical things [11]. Users with disabilities are present in several scenarios. The user-centered approach is especially useful to create design ideas in the field of accessibility [12]. It would not be possible to include every possible AI assisted musical device in a given scenario. Our objective is just to explore some possibilities that cover the different device classifications proposed in our taxonomy.

| | |
|---|---|
| **1. Musical instruments.** | (a) AI assisted instruments. |
| | (b) Augmented instruments. |
| **2. Music processors.** | (a) Instrumental modifiers. |
| | (b) Voice modifiers. |
| | (c) General sound processors. |
| **3. Music generators.** | (a) Instrumental. |
| | (b) Voice. |
| | (c) Combined. |
| **4. Music recommendation devices.** | (a) Ambient aware recommendation. |
| | (b) User aware recommendation. |
| | (c) Combined. |
| **5. Music-related feedback systems.** | (a) Personal Feedback. |
| | (b) Ambient Feedback. |
| | (c) Combined. |
| **6. Educational Aids .** | (a) Music Education. |
| | (b) General educational support. |
| | (c) Rehabilitation. |

**Table 1.** AI-assisted Musical dEvice (AIME) Taxonomy

2.2.1. Instrument scenarios

The area of intelligent musical instruments [13] includes an important subset of musical devices and has a wide range of applications that we will present in four example scenarios.

Able Instrument Scenario

Mike had an accident that led to a problem that prevents him from playing with his right hand. However, he would like to continue playing the bass in a small blues band. Mike thought he would not be able to play again as a bass player, as most instruments require significant ability with both hands. There are several alternatives to adapt the instrument to his physical capabilities [14], but finally he settled on a small robotic mechanism that can detect which string is he fretting with his left hand and pluck it. This device can hear what other members of the band are playing and dynamically adapt to the tempo and genre of the song by varying the rhythms and patterns it plays.

Although the results do not match his earlier performances, Mike is still able to play well enough and have fun with his friends' band.

Drum Stroke Scenario

Toby recently had a stroke that left him with reduced mobility in his right hand. In his rehabilitation clinic, they proposed that he should follow complementary music-supported therapy (MST) in which he controls a set of midi drums through his hand gestures [15] which are detected through electromyography signals (EMG). The drums can play almost autonomously at the beginning of therapy and allow control of an increased number of variables as Toby progresses in his recovery.

The rehabilitation device keeps track of Toby's progress and periodically sends reports to his therapist. When Toby goes to the clinic for an in-person session, the therapist will discuss his progress and adapt the MST accordingly.

Teach and Play Scenario

Mary wants to start playing the concertina and is following a well-known book and taking some lessons online. However, she does not like the sound that she is producing with the instrument at the moment and refuses to play it anywhere. A friend tells her about Inteltina, an intelligent didactical concertina that augments Mary's abilities and helps her produce a nice sound. The instrument assistance dynamically decreases as Mary's playing capabilities improve.

Although Mary plays reasonably well with Inteltina, her online teacher warns her that this type of instrument sometimes backfires as the student becomes lazy and her abilities stagnate [13].

TherAImin

Sara is a computer scientist who plays piano as a hobby. Recently she has become fascinated by the discovery of the Theremin [16]. Figure 1 shows an early implementation of Theremin. Being an AI specialist, she believes that the design can be clearly improved with the help of AI. Thus, she decides to become a 'digital luthier' and to create a new instrument that is faithful to the original Theremin concept. The TherAImin keeps the pitch and volume antennas of the original instrument, but includes an AI-based gesture recognizer to change the timbre of the instrument [17] according to hand gestures.

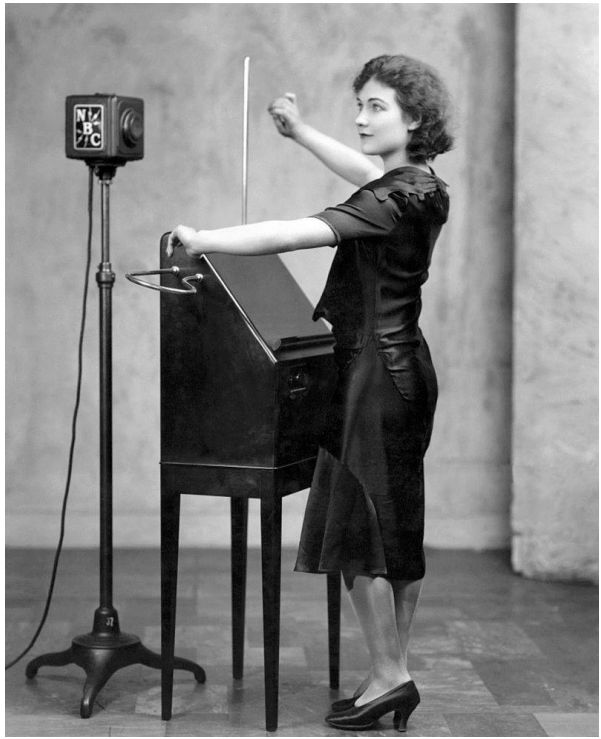

**Figure 1.** Alexandra Stepanoff playing the theremin, 1930

This scenario reflects the creation of new digital AI-supported musical instruments. Several interesting reflections on this topic can be found in [3].

This type of instruments are fun to build and play, but it can be difficult to create a community of users around them.

2.2.2. Audio processing Scenarios

This area includes instrument processors, voice processors, and generic audio processors.

Boogie Boogie Scenario

Saul is a professional guitar player. He would love to have a Mesa Boogie Mark V amplifier but the price is too high for him. Saul knows that there are emulations for this amp for several Digital Audio Workstations (DAWs) including Cubase which he regularly uses. However, Saul would like to have the emulation as a pedal he can easily carry. He has several friends who work in a small start-up company that designs embedded deep learning devices and learns from them that the boogie can be emulated by an AI system [18] that can be implemented using a Coral Edge TPU accelerator [19].

In a few months, Saul has tested the device and the company is starting to sell the BoogieBoogie Pedal.

DeepTuner Scenario

Sara is a singer who regularly uses a pitch-correction voice processor for her performances. Currently, she uses an AI enhanced version of Antares Auto-Tune [20] on an Avid Carbon Device. She is satisfied with the natural feeling, and virtually unnoticeable delay that this hardware/software implementation bring to her performances. Nevertheless, she would love a similar pitch correction implementation in a smaller and cheaper device [21].

DeepAFx Scenario

Kyra is a production Engineer. Since she discovered the Deep Learning-based LV2 DeepAFx plug-in framework [22] she regularly uses it to control her DAW and to introduce several effects. Although she always fine-tunes the work manually, the use of the framework has clearly improved her schedule. Kyra would love to have a device with an embedded version of these plug-ins for live performances.

### 2.2.3. Music Generator Scenarios

In this subsection we present two scenarios that rely on the use of different AI based music generators.

#### On Hold Scenario

Peter has a small online seller business with a telephone customer service line. He wants some copyright-free music to keep the costumer on hold while an agent can handle their call. He wants the music to change according to the expected waiting time, the time of the day, and other circumstances.

Peter has heard about AI-based music generation technology [5] and after searching online decides to select some compositions made using AIVA [23] and computoser [24]. Peter consults with his guitar player friend Saul to help him decide which parameters would be best for the different musics that he wants for the costumer service line. An automated controller dynamically changes the generator parameters to create the desired result.

Peter would like to be able to estimate the emotional state of the client [7] and change the music accordingly; however, this is not possible in a standard phone call. When clients use the costumer service app, the music changes according to their comments [25]. All the generators in this scenario produce symbolic music in midi format. This format is suitable for instrumental music and produces results of a quality that can be adequate for the proposed scenario.

#### Singing Elevator Scenario

Mia is a Design Engineer for a large elevator company. In their latest models, the elevators are fitted with a screen that mainly provides news and weather information. Mia wants to have copyright-free background songs while the elevator is in use.

After studying several alternatives, Mia decides to generate the songs dynamically based on the characteristics of the building (residential, commercial, neighborhood, etc.). To generate the songs, she uses the OpenAi Jukebox generator [6] and updates the sons on a regular basis. The entire selection of songs according to the different situations is performed by the elevator media controller which can also be considered a musical thing. This scenario uses a nonsymbolic direct audio music generator. This type of generator is much less common than the symbolic alternatives, but the results are becoming acceptable by final users in the last years.

### 2.2.4. Music Recommendation Device Scenarios

#### Emotiwatch Scenario

Sam is a sports and music fan. Every morning he runs for an hour. While running, Sam likes to listen to music. His musical choices clearly depend on his mood. For years Sam has selected his songs directly, but he would prefer, at least sometimes, that his smartwatch would do the selection for him. It is well known [26,27] that emotional states and stress can be predicted using AI technology from physiological indicators. These are mainly electro-dermal activity (EDA), heart rate variability (HRV), and, to a smaller extent, peripheral oxigen saturation (Spo2). Several wearable devices, including smartwatches such as Fitbit charge 2 or Sense [28] or research-oriented Empatica E4 wristband [29] are capable of measuring at least a subset of these parameters. Sam finds an app for his watch [30] that selects music based on his mood. The watch, which was already a musical thing, becomes an AI-assisted musical device and lets Sam keep his mind on running.

#### iClock Scenario

Jane, like a great part of the population in many countries has been having lack of sleep problems for a long time. The relationships between sleep disorders and anxiety, depression, overweight, and diabetes are well known by the medical community [31]. As part of her treatment, her psychologist tells Jane that some new devices could possibly help restore her sleep quality. Among these devices, Jane finds iClock, a new device that monitors her sleep, using Jane's smartwatch, and modifies her rise up routines taking into account her schedule needs, the sleep monitoring data and an estimation of her emotional state. Among the different aspects that iClock controls is the selection and modification of the melodies according to the selected rising up routine. Thus, iClock is, among other things, an AI assisted musical device,

Following her therapist recommendations, including the use of iClock, Jane's sleep patterns improve, which in turn is clearly reflected in an improvement of her quality of life.

2.2.5. Feedback device Scenarios

RumbleRumble Scenario

Gina has a moderate hearing problem. She likes to go to concerts with friends. However, she feels that she is loosing an important part of the information. Recently she learned about the existence of the subpac backpack [32] that uses haptics, interoception, and bone conduction to deliver bass sensation to even profoundly deaf users. Although the current version of the device requires an external computer to run the software, Gina is using an experimental version that runs in an embedded controller, thus, making the subpac a personal feedback AIME.

Let there be light scenario

Nico really likes to go to rock music performances. He especially loves when people start following music with their lighters. In some recent concerts, this has even improved due to new musical device technology. When Nico went to his last concert, he was given a PixMob led wristband. These devices have a set of pre-programed effects that are trigered usually by a human operator.
Nevertheles, the possibility of an AI based controller that decides which effect to apply according to both the concert and the carrier circumstances is currently perfectly feasible. In this way, the wristband will become an AIME.

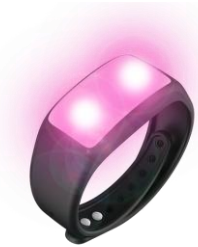 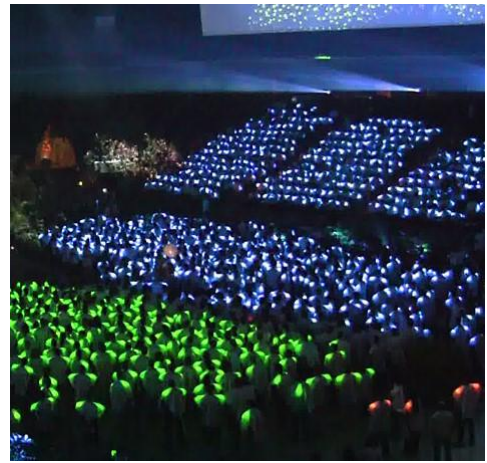

**Figure 2.** LED based ambient feedback devices

2.2.6. Educational Scenarios

Teach and Play Scenario - Again

The Teach and Play Scenario presented in Subsection 2.2.1.3 is also clearly an educational Musical Device scenario and could have been presented in this subsection as well.

The magiFlute Scenario

John is 13 years old and has a moderate learning disability. His music teacher recommends that he use a new accessible digital musical instrument (ADMI) [33] known as the MagiFlute. This instrument is an Electronic Wind Instrument (EWI) [34], which is similar to a recorder, but does not produce sound directly. It has sensors for wind and touch pressure and controls a synthesizer through an embedded deep learning system. It also uses John's iPad to help him remember what to play and how to play. It even has the possibility of automatically correcting what John is playing when configured in this way. With the magiFlute, John participates in the school band and is becoming a better standard recorder player every day.

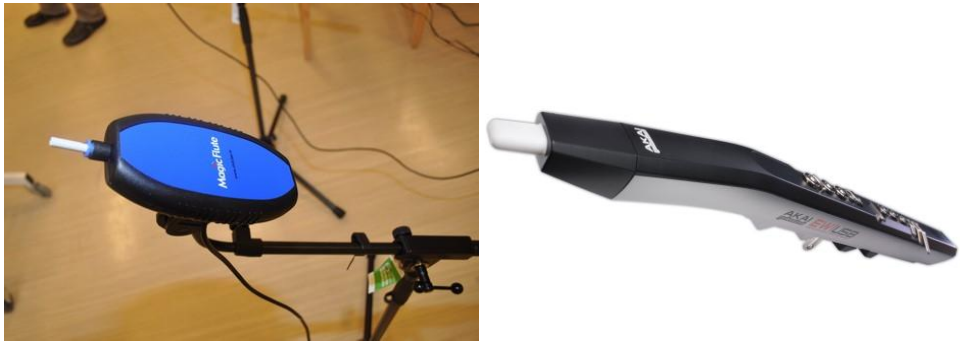

**Figure 3.** Magic Flute and typical EWI

Magic Flute Scenario

In our last scenario, we use the term magiFlute for our proposed instrument, as its housemate, the 'Magic Flute', is a completely different existing ADMI which is an EWI that is controlled by very small head movements [35]. This instrument is played by Ellen, who has a spinal cord injury as a result of a motorcycle accident.

*2.3. Processing architecture*

In this work, we propose a generic architecture for the design of musical devices. This architecture is based on a multilayered approach. The proposed layers are structured as follows:

1. User stimuli capture and processing layer.
2. Embedded learning Layer
3. Music adaptation layer 4. Music production layer
5. User feedback layer.

The block diagram for this proposed architecture is shown in Figure 4.

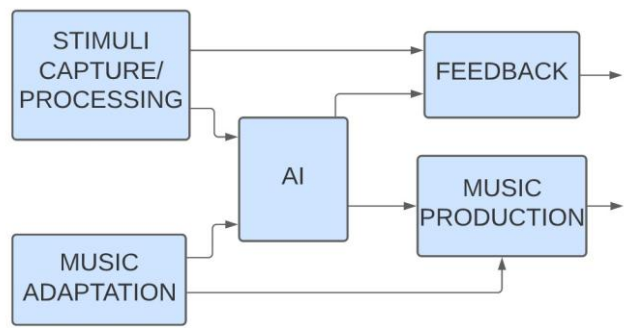

**Figure 4.** Generic AIME Block diagram

After presenting the methodology used to test the theories discussed in the Introduction, the results will be detailed in the next section.

**3. Results**

In this section, we present a possible implementation for some of the devices proposed in the scenarios. In this way, we will verify the suitability of the proposed generic architecture and, thus, the usefulness of the framework presented.

*3.1. Scenario Implementation*

We will briefly describe possible implementations of TherAImin. This implementation is presented to show that the proposed framework provides a usable foundation for building AI-assisted devices and describing them in a systematic manner.

We think that even though we do not present a device for each of the possible categories, the difference between the selected AIMEs is wide enough to show that, in principle, any AIME can be implemented using the framework.

### 3.1.1. TherAImin

As discussed in section 2.2.1 the Theremin is an instrument with two antennas that is controlled by the player without touch interaction. The block diagram of the Theremin is shown in Figure 5. The TherAIMin is an AI-assisted variation of the original instrument, where hand gestures are used to control the timbre.

Although we could have implemented TherAImin without antennas using, e.g. Mediapipe Handpose [36] we have decided to be more faithful to the original instrument and thus use the [37] which provides a versatile Theremin implementation with Pitch and Volume outputs.

Thus, openTheremin antennas act as part of the user-stimulus capture layer. The other part of this layer is a camera that is used to capture the user's hand gesture.

We will interface openTheremin using a Raspberry Pi board with an RPI-GP90 pulse signal IO hat. This is part of the stimulus adaptation layer. The other part of this layer is made up of the video interface already available in the raspberry pi.

The embedded learning layer is built using Google's Teachable Machine [38] accelerated with a Coral Edge TPU accelerator. The approach is very similar to [39] where a machine that can be trained is used to recognize objects. With this approach, the accelerated embedded system classifies the gestures in the number of trained classes. It is important to keep the gesture classes different, and it is also essential to train a wide class of gestures and other images that the camera may see in the background class [40]. An advantage of the therAImin is that when the AI system makes a wrong decision, this
will affect the timbre and the effects, but not the volume and the pitch.

The sound production layer is implemented on raspberry pi using sonic PI [41]. The selection of sound pitch and volume is done by a small Processing program that produces OSC [42]. Open Sound Control (OSC) is a protocol to connect sound synthesizers, computers, and other multimedia devices for purposes such as musical performance or show control. Many music-related software tools, including sonic PI support the OSC protocol. The OSC protocol uses UDP (or TCP) packets and thus can run either in a single embedded system or be distributed over a network.

The selection of timbre and effects is done by the learning layer as a function of the gestures and sent to the generation layer using OSC. In this way, TherAImin produces audio as a function of the hand positions captured by the antennas and the gestures captured by the camera and recognized by the teachable machine.

The TherAImin, as an extension of the Theremin, can be considered in the augmented musical instrument class in the FAIME taxonomy.

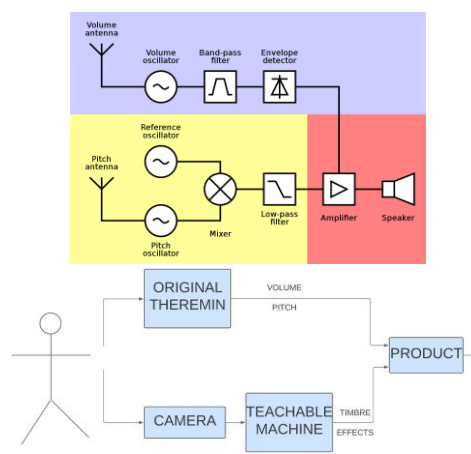

**Figure 5.** Block diagram of the Theremin ad the TherAImin.

## 4. Discussion

The approach used to implement TherAImin could be used for simpler and more complex devices. As an example, the 'Singing Elevator' can uses the presence, temperature, humidity and noise sensors and additional information from the Internet as user stimuli. Using a simple learning layer (local or not), it will decide from which category it should retrieve the generated music. The production layer
would be a simple player with possible adaptations to handle noise in the cabin or other issues.

As a further example, the emotiWatch uses the wearable device sensors as a stimuli layer, preprocesses them in the stimuli adaptation layer, estimates the user's emotional state, and selects the music in the learning layer and outputs the music through a player in the production layer. The same approach can be used to start the design of any AIME.

It is clear that many other approaches could have been proposed, but FAIME is simple and gives clear insights into the musical device design process.

## 5. Conclusions

In this work, we have presented a useful framework for the classification, understanding, and design of AI-assisted musical devices. We have shown a very wide range of devices that can fall into this category including such different things as accessible instruments for disabled musicians or alarm clocks to help people with sleeping disorders.

We have presented a quite detailed implementation of a variation of a successful musical instrument designed in the 1920s, the Theremin. Our augmentation allows the player to select timbre and effects in real time through hand gestures but also helps to keep the look and feel of the original instrument if it is played with open hands.

We also included a short description of the design of other AIMEs to show the usefulness of the framework.

In future work, we will evaluate the user experience of TherAImin with musicians and study possible modifications for performers with disabilities using the powerful embedded intelligent system.

**Author Contributions:** Conceptualization: M.C, L.M-S., F.C.M, MJ. E. and A.C.; methodology: M.C., MJ. E. and
A.C.; and A.C.; formal analysis: F.C.M, MJ. E. and A.C.; investigation: M.C., L.M-S. and A.C.; writing:M.C, L.M-S., F.C.M and A.C.; supervision: F.C.M, MJ. E. and A.C. All authors have read and agreed to the published version of the manuscript.

**Funding:** This work was supported by the NICO project (PID2019-105455GB-C31) from Ministerio de Ciencia, Innovación y Universidades (Spanish Government) and by the DAFNE (US-1381619) Consejería de Economía y Conocimiento (Junta de Andalucia).

**Conflicts of Interest:** The authors declare no conflict of interest.

## Abbreviations

The following abbreviations are used in this manuscript:

| | |
|---|---|
| ADMI | accessible digital musical instrument |
| AIME | AI assisted Musical dEvice |
| AI | Artificial Intelligence |
| DAW | Digital Audio Worksation |
| DL | Deep Learning |
| DTL | Deep Transfer Learning EDA |
| | Electro Dermal Activity |
| EWI | Electronic Wind Instrument |
| HRV | Heart Rate Variability |
| IME | Intelligent Musical dEvice |
| IoMusT | Internet of Musical Things |
| MIDI | Musical Instrument Digital Interface |
| ML | Machine Learning |
| MST | Music Supported Therapy MusT |
| | Musical Thing |
| OSC | Open Sound Control |
| SpO2 | Peripheral Oxigen Saturation |